\documentclass[journal]{IEEEtran}
\usepackage[T1]{fontenc}
\usepackage{commath}
\usepackage[space]{grffile}
\usepackage{times}
\usepackage{soul}
\usepackage{url}
\usepackage[colorlinks,linkcolor={blue},pagebackref=true,breaklinks=true,colorlinks,bookmarks=false]{hyperref}
\usepackage[utf8]{inputenc}
\usepackage[small]{caption}
\usepackage{graphicx}
\usepackage{amsmath}
\usepackage{amssymb}
\usepackage{xspace}
\usepackage{booktabs}
\usepackage{stmaryrd}
\usepackage{color}
\usepackage[export]{adjustbox}
\usepackage{xspace}
\usepackage{multirow}
\usepackage{csquotes}
\usepackage{textcomp}

\usepackage{mathtools}
\usepackage{hyperref}
\usepackage{fontawesome}
\usepackage{enumitem}
\usepackage{algorithmic}

\usepackage{url}
\usepackage{graphicx}
\usepackage{subfigure}
\usepackage{makecell}
\usepackage{soul}
\usepackage[linesnumbered,ruled,vlined]{algorithm2e}
\usepackage[edges]{forest}
\usetikzlibrary{shadows.blur}

\usepackage{amssymb,bm}
\usepackage{amsmath}

\usepackage[normalem]{ulem}
\usepackage{xspace}
\newcommand{\latinphrase}[1]{\textit{#1}} 
\newcommand{\etal}{\latinphrase{et~al.}\xspace}

\usepackage[utf8]{inputenc}
\begin{document}

\title{A Comprehensive Overview of Large Language Models (LLMs) for Cyber Defences: Opportunities and Directions}

\author{Mohammed Hassanin, Nour Moustafa
\thanks{M. Hassanin is with University of South Australia,  Australia, and the Faculty of Computers and Information, Fayoum University, Egypt.}
\thanks{N. Moustafa is with University of New South Wales,  Australia.}
% \thanks{S. Anwar is with The Australian National University, Data61-CSIRO, The University of Technology Sydney, and the University of Canberra, Australia.} 
% \thanks{I. Radwan is with the University of Canberra, Australia. }
% \thanks{Fahad S. Khan is with Mohammad Bin Zayed University of Artificial Intelligence, Abu Dhabi, UAE, and Computer Vision Laboratory, Linkoping University, Sweden.} 
% \thanks{A. Mian is with The University of Western Australia, Australia.}
% \thanks{Corresponding Author: Saeed Anwar. Emails: saeed.anwar@anu.edu.au}% 
}

\maketitle

\begin{abstract}
The recent progression of Large Language Models (LLMs) has witnessed great success in the fields of data-cetric applications. LLMs trained on massive textual datasets showed ability to encode not only context but also ability to provide powerful comprehension to downstream tasks. Interestingly, Generative Pre-trained Transformers utilised this ability to bring AI a step closer to human being replacement in at least data-centric applications. Such power can be leveraged to identify anomalies of cyber threats, enhance incident response, and automate routine security operations.
We provide an overview for the recent activities of LLMs in cyber defense sections, as well as categorization for the cyber defense sections such as threat intelligence, vulnerability assessment, network security, privacy preserving, awareness and training, automation, and ethical guidelines. Fundamental concepts of the progression of LLMs from Transformers, Pre-trained Transformers, and GPT is presented. Next, the recent works of each section is surveyed with the related strengths and weaknesses. A special section about the challenges and directions of LLMs in cyber security is provided. Finally, possible future research directions for benefiting from LLMs in cyber security is discussed.
\end{abstract}

\IEEEpeerreviewmaketitle

\section{Introduction}
\IEEEPARstart{C}yber security is an essential realm that focuses primarily on safeguarding digital systems (computer systems, networks, and data) from any wrongdoings, including unauthorized access or interruption. While technology has  increasingly been integrating into every facet of life, the necessity of cyber security is growing as well. It encompasses a vast set of practices, ranging from guidelines to applications designed to protect data. With the development of digital systems including the Internet of Things (IoT) \cite{moustafa2022dfsat}, Internet of Vehicles (IoV) \cite{contreras2017internet}, Internet of Battlefield Things (IoBT) \cite{zhu2021invisible}, and Cyber-Physical Systems (CPS) \cite{baheti2011cyber}, as well as technologies such as smartphones, tablets \cite{merle2015computers}, autonomous vehicles, digital twins \cite{juarez2021digital}, cryptocurrencies \cite{sai2019privacy}, and smart homes \cite{farooq2021iot}, the scope of cyber security is expanding accordingly. The more these technologies integrate into daily life, they also widen the potential landscape for cyber threats. Therefore, they necessitate advanced security technologies to safeguard such systems from intrusions and inaccessibility. 

Cyber defense is a realm of cyber security that is responsible for detecting, preventing, and responding to attacks or cyber threats \cite{oyelami2020cyber}. It combines different advanced technologies such as deep learning (DL) \cite{salim2023deep} and large language models (LLMs), as well as maintaining set practices such as threat intelligence and incident response in order to achieve security to the digital system. The goal of cyber defence is twofold: to safeguard digital systems from cyber threats and to build robust systems able to recover from any attacks. Lately, cyber attacks has become more sophisticated, that dictates drawing intensive attention to build proactive technologies for cyber defence. To achieve that, research community need to spend more time and effort to be ahead of cyber attacks techniques.

In this digital era, the necessity of cyber defense in cannot be overstated. As every domain of the life such as economy, health, education, communities, and social welfare, the need to protect such systems from cyber threats becomes mandatory. Ranging from data breaches and ransomware to espionage, the scope of cyber threats dictates the urgent need for robust cyber defense strategies \cite{sviatun2021combating}. For instance, Sviatun \etal reported that cyber crimes cost US around 1 trillion every year and 20 billion for the whole world \cite{shukla2024security}, highlighting the economic impact of lacking cyber defenses. Moreover, such development of cyber defense techniques also supports the protection of individual privacy. Therefore, cyber defense is not only an integral necessity, but it is a fundamental component of daily life to ensure a secure, stable, and fair society.

\begin{figure}[] 
\centering 
\includegraphics[width=\linewidth]{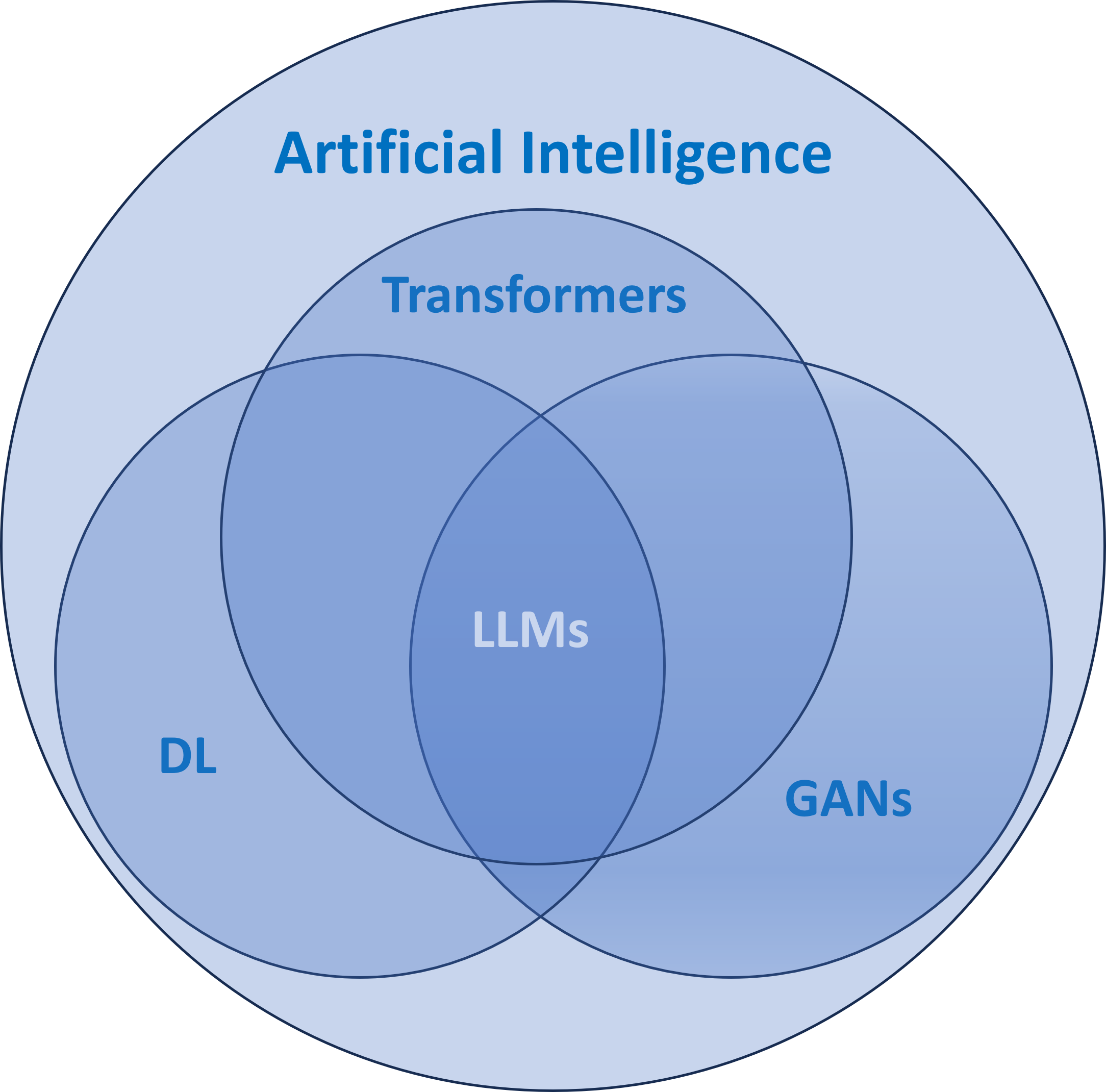}
   \caption{The evolution of LLMs and GPT is based on deep learning, GANs, and Transformers.}
  \label{fig:LLMs_evolution}
\end{figure}

Several strategies can be used for cyber defence, however, machine learning stays on top of all the technologies. As the recent progression of machine learning and large language models (LLMs) represents a paradigm shift in AI, it impact all the science branches, particularly cyber security. Machine learning algorithms used small datasets to learn simple computer vision tasks such as plate numer recognition and other simple tasks. The invention of Deep Learning along with the increase in computational power was the game changer to machine learning paradigm, for it resolved the challenges of neural network such as that learning a networks with many layers\cite{hassanin2024visual}. Meanwhile, Natural Language Processing (NLP) utilised this progression of deep learning and achieved paradigm shift in text processing by the invention of Transformers, which in turn impact the whole machine learning field \cite{vaswani2017attention}. It  achieved that success by allowing the model to weigh the importance of all parts of the input space simultaneously. This simultaneous processing helped encode long-range dependencies and thus the context \cite{hassanin2022crossformer}. Parallel to these advancements, Generative Adversarial Networks (GANs) were first proposed by Goodfellow \etal as a different way of training that instead based on data generation models \cite{goodfellow2014generative}. Simply put, it composes two neural networks: a generator and a discriminator, which are trained concurrently using adversarial techniques. The generator produces synthetic data that closely resemble to the orginal sample, while the discriminator evaluates how far the generated image from the original one. They have been used in many designs and diverse fields including image synthesis, style transfer, and advancements in medical imaging techniques \cite{creswell2018generative}. These three different designs of deep learning, \textit{i.e., deep neural networks, GANs, Transformers} paved the way to LLMs evolution as Figure \ref{fig:LLMs_evolution} shows.

\begin{figure*}[] 
\centering 
\includegraphics[width=\linewidth]{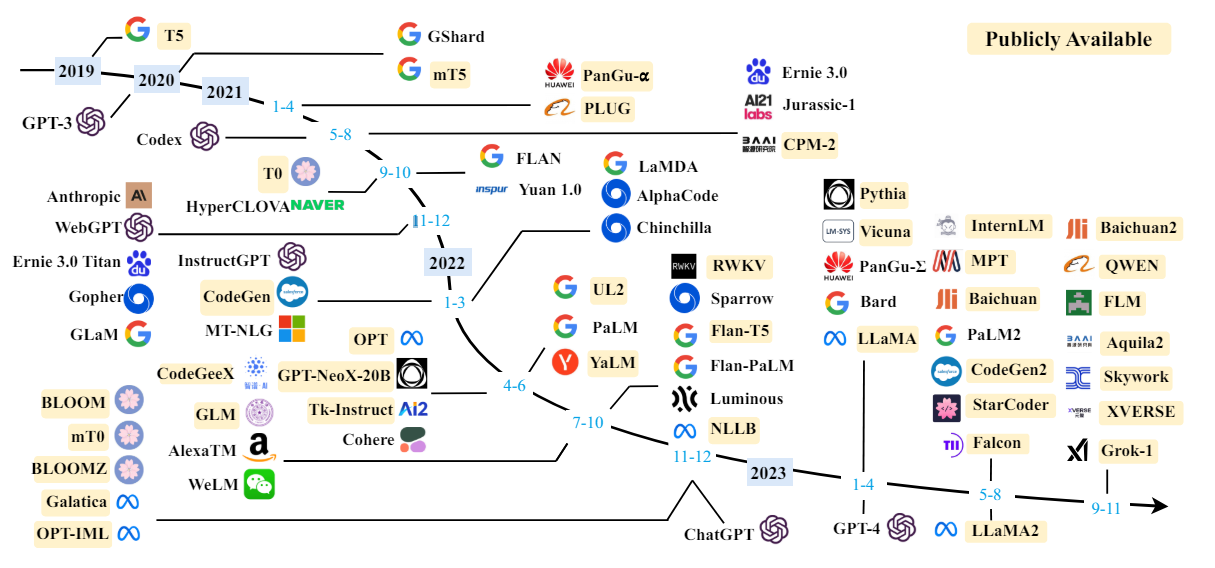}
   \caption{The publicly available LLMs in the most existing in recent years in  a timeline. The timeline is ordered based on the publishing date. Figure from \cite{zhao2023survey}}
  \label{fig:LLMs}
\end{figure*}

\begin{figure}[] 
\centering 
\includegraphics[width=\columnwidth]{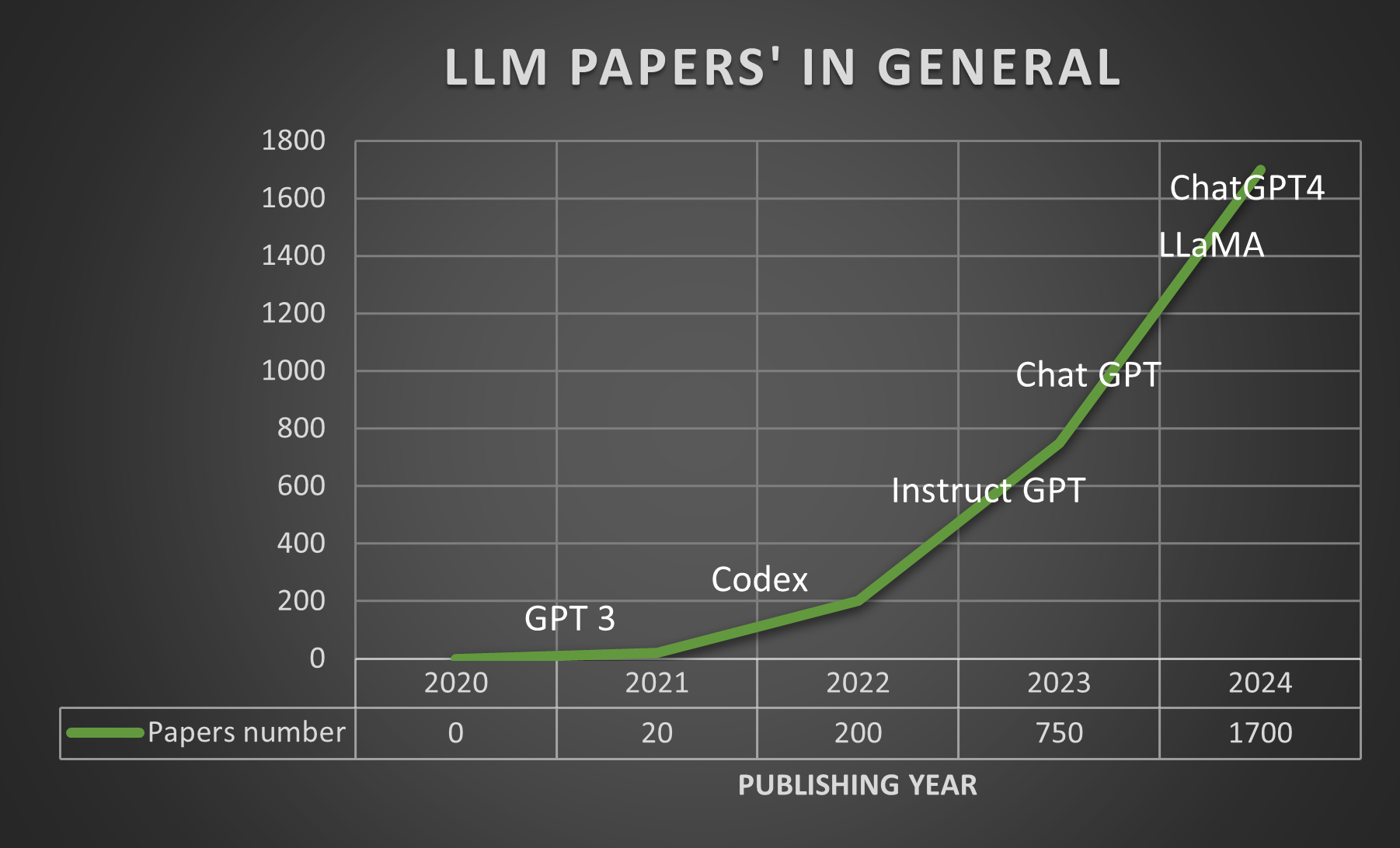}
   \caption{Visual description of the number of LLMs usage in research in all fields in general.}
  \label{fig:general_numbers}
\end{figure}

\begin{figure}[h] 
\centering 
\includegraphics[width=\columnwidth]{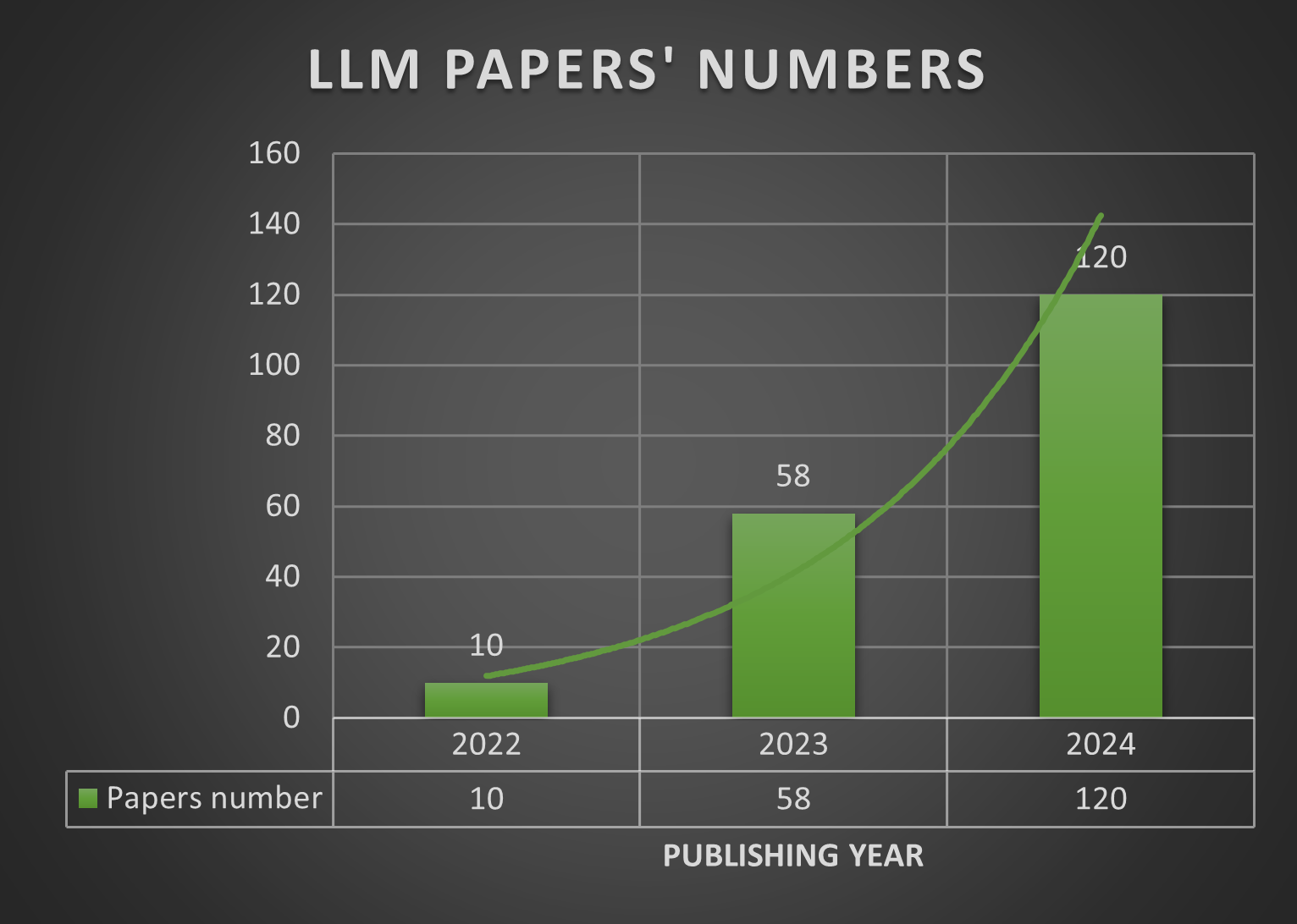}
   \caption{Visual description of the number of LLMs usage in research in all fields in cyber security fields.}
  \label{fig:papers_number}
\end{figure}

As a new step towards the current progression of LLMs,  Bidirectional Encoder Representations from Transformers (BERT) has been introduced with high rates of success. Such models are trained on massive corpora of text data and then they generate coherent. Contextually, these models generate appropriate text that can successfully perform language translation and answer questions with high accuracy \cite{devlin2018bert, brown2020language}. Recently, Generative Pre-trained Transformers (GPT) have been proposed as a hybrid evolution of Generative Neural Networks (GANs) and Masked Pre-trained Transformers.
The result of this blend is that models are capable of generating coherent and contextually relevant text given a good input, called a prompt. Thus, a significant step forward towards more intelligent AI that will help automating a lot tasks, including text comprehension and visual understanding \cite{saha2023llm}. One of the beneficiaries of this progression is certainly cyber security in both sides; threats and defence \cite{dong2024attacks}. Large Language Models (LLMs) like GPT (Generative Pre-trained Transformers) are becoming increasingly significant in the field of cyber defense, offering a range of capabilities that enhance both the effectiveness and efficiency of cybersecurity operations. Their extensive training on diverse datasets enables them to understand and generate human-like text (see Figure \ref{fig:LLMs}), which can be leveraged in several critical areas of cyber defense. The illustrated power of LLMs in analyzing massive amounts of data, including web logs, network traffic, and regular datasets, will certainly push the cyber defense realm ahead, particularly with early threat detection, which as a result allows faster and more timely response \cite{karlzen2023automatic}. Such capability of text comprehension and contextual understanding is required for cyber security routinely tasks such as documentation and email responses, which will free security experts to focus on complicated threats. Another aspect is the benefits of LLMs in generating training scenarios and simulations which will enrich the training of algorithms as well as the staff. Even the realm of governance and guidelines, LLMs provide an amazing features to cyber security by helping in keeping regulations up to date (see Figures \ref{fig:general_numbers} and \ref{fig:papers_number}).

\begin{figure*}[h] 
\centering 
\includegraphics[width=.9\linewidth]{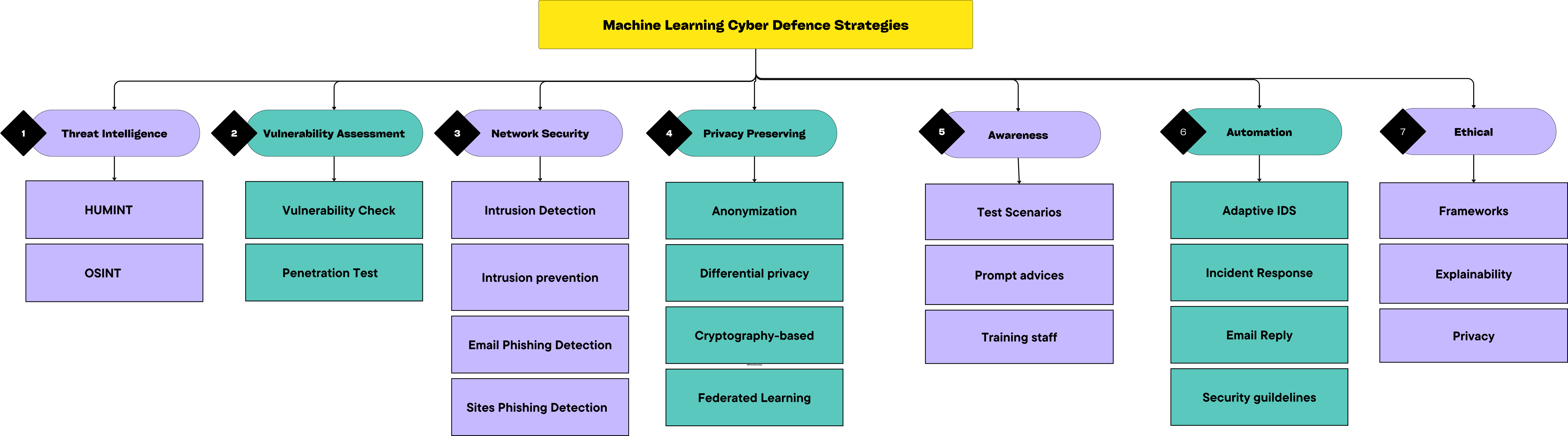}
   \caption{A taxonomy of using LLMs in cyber defence sections. They are categorized based on the type of security. Some techniques may intersect in multiple categories; in this case, they are grouped based on the most dominant characteristic.}
  \label{fig:taxonomy}
\end{figure*}

In this survey, we review LLM-based cyber defence techniques specific to machine learning defense strategies. We cover the most of the unique approaches in the cyber defense since the evolution of LLMs. Our survey broadly classifies LLM techniques into threat intelligence, vulnerability assessment, network traffic security, privacy preservation, personnel awareness, ethical security. We noted that some of the techniques fit within more than one category; however, we put them under the dominant of the method. Such a classification helps the researcher to draw attention to the gaps, as well as easiness to get insights. Figure~\ref{fig:taxonomy} shows the categories of LLM-base cyber defence technologies.

We highlight that this review is focusing only on cyber defence due to vastness of cyber security fields shown in  Figure~\ref{fig:papers_number}. It is clear from Figure~\ref{fig:general_numbers} that the number of published papers in the last year has dramatically increased compared to before as we anticipate to even exponential increase the upcoming years. Further, our article lists the main articles with unique ideas to provide the cyber defence community with keypoints to find the proper solutions for the real gaps and issues. Research gaps are  provided in the current research context, as well as plausible research directions along with with future areas are projected and discusses.

Since LLMs have been used across cyber security,  few surveys \cite{zhang2024llms, li2024personal} review the recent trends of LLMs in cyber security. Although LLMs provide significant accuracy, this comes at the account of a lot of issues, which limit their feasibility real-life domains. Cyber-defence strategies need consistent adaptation to fix the ever-changing attacks techniques and tricks. In \cite{yao2024survey}, Yao \etal provide a survey on all the cyber security rather than the defense. Also, there is a recent review published on drawbacks and vulnerability of LLMs ~\cite{wu2024new}, which different from our scope. In contrast, our article is primarily focusing on the LLM-based cyber defence strategies and their impact on the light of the previous works.

\section{Threat Intelligence}
Threat intelligence is a realm of cyber defence that specialises on network traffic engineering including collection, evaluation, and analysis of any potential or current threats that harms the system \cite{conti2018cyber}. It focuses primarily on understanding the cyber attacks and the environment of the cyber criminal or attacker. The goal is to use all the techniques to imagine the cyber criminal way of attack to respond with the best strategy and even expect it before hand. By collecting and analyzing such data from various sources, including internal technical programs,  human intelligence (HUMINT), open-source intelligence (OSINT), social media, etc, cyber threats landscape can be understood and predicted. The result of this process will be used to update security strategies, help incident response teams, as well as refine threat detection capabilities, and thus improve the whole security operations. 

\begin{figure*}[th!] 
\centering 
\includegraphics[width=.9\linewidth]{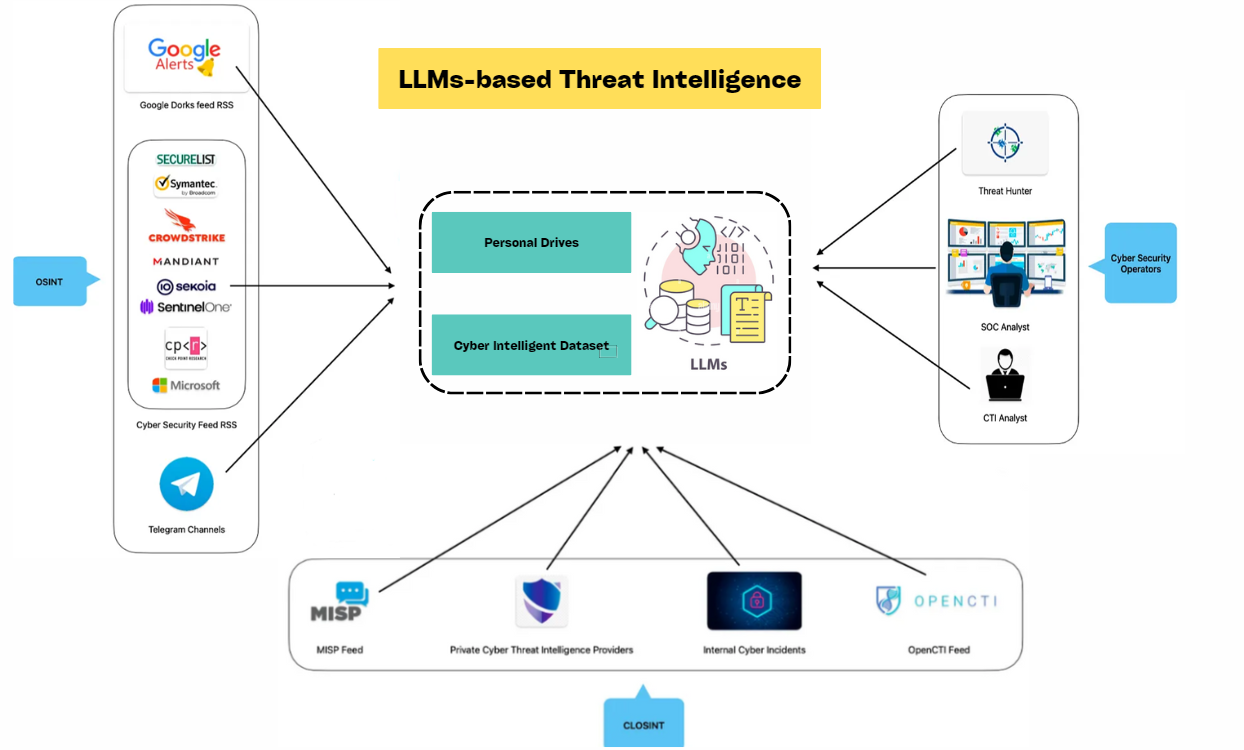}
   \caption{Visual description of the LLMs-based threat intelligence steps. LLMs is used to analyze the possible exploits and threats.}
  \label{fig:threat_intelligence}
\end{figure*}
Successful threat intelligence outputs multiple operations: collection of important data,  extracting actionable insights through analysis, updating the appropriate stakeholders of the results, and finally list these insights in the security operation tasks. The outcomes of threat intelligence is helpful for organisation to have the best response to the attacks. This is the first step process of attack response, however, it is complicated due to the advancement of the attacks and the tricks, which dictates more attention from the research community of cyber defence.  In this section, we are reviewing LLM-based penetration testing strategies as we provide discussion on each method, significance and drawbacks (see Figure \ref{fig:threat_intelligence}).

Sufi \etal used GPT models to improve the capabilities of OSINT to automatically extract, analyse, and summarise important information from historical cyber incident reports \cite{sufi2024innovative}. They succeeded in improving the accuracy of threat intelligence and as a result ability of forecasting future cyber threats.  This work emphasised LLMs capacity to conclude deep insights to cyber security professionals that help in answering efficiently to emergent risks. Siracusano \etal used LLMs to address the issue of extracting pertinent information from unorganised cyber threat intelligence (CTI) data, which is usually a time-consuming and labor-intensive process \cite{siracusano2023time}. They presented a dataset of 204 real-world reports that adhere to the STIX ontology, which is considerably larger than existing datasets. Then, GPT-3.5 is used to enhance the effectiveness of CTI extraction. The empirical experiments demonstrate that aCTIon surpasses the previous methods with a significant margin in identifying items such as malware, threat actors, and attack patterns from cyber threat data. In \cite{mitra2024localintel}, LOCALINTEL introduced a step forward in the process of generating organization-specific threat intelligence automation using LLMs. Hu \etal used LLMs to create comprehensive knowledge graphs, notably LLM-TikG extract and organising threat intelligence from extensive amounts of unorganised data \cite{hu2023llm}. This allows for more accurate anticipation, and thus, more mitigation of cyber attacks. Utilising LLMs improved the accuracy and speed of knowledge graph generation that provide a strong framework to promptly detect and address emerging threats. Sewak \etal presented a novel method combining LLMs and semantic hypernetworks to improve the detection, analysis, and response to cyber threats efficiently \cite{sewak2023crush}. This hybridization is proposed to enhance the identification of advanced cyber threats along with flexible solution that can be adjusted to different cyber security obstacles. Hays \etal incorporated LLMs into threat intelligence exchange to replicate intricate and authentic security situations which enable users to participate in decision making \cite{hays2024using}. The development of exercises that utilise LLMs to create a wide range of threat scenarios and tactics for response have been explored. Moreover, it evaluates the consequences of LLM-enhanced TTXs on the training of security personnel, the improvement of collaborative problem-solving. It provided recommendations for LLMs in security TTXs to take into account technological and ethical considerations to optimise their effectiveness along with responsible behaviour.

\begin{figure}[th!] 
\centering 
\includegraphics[width=.9\linewidth]{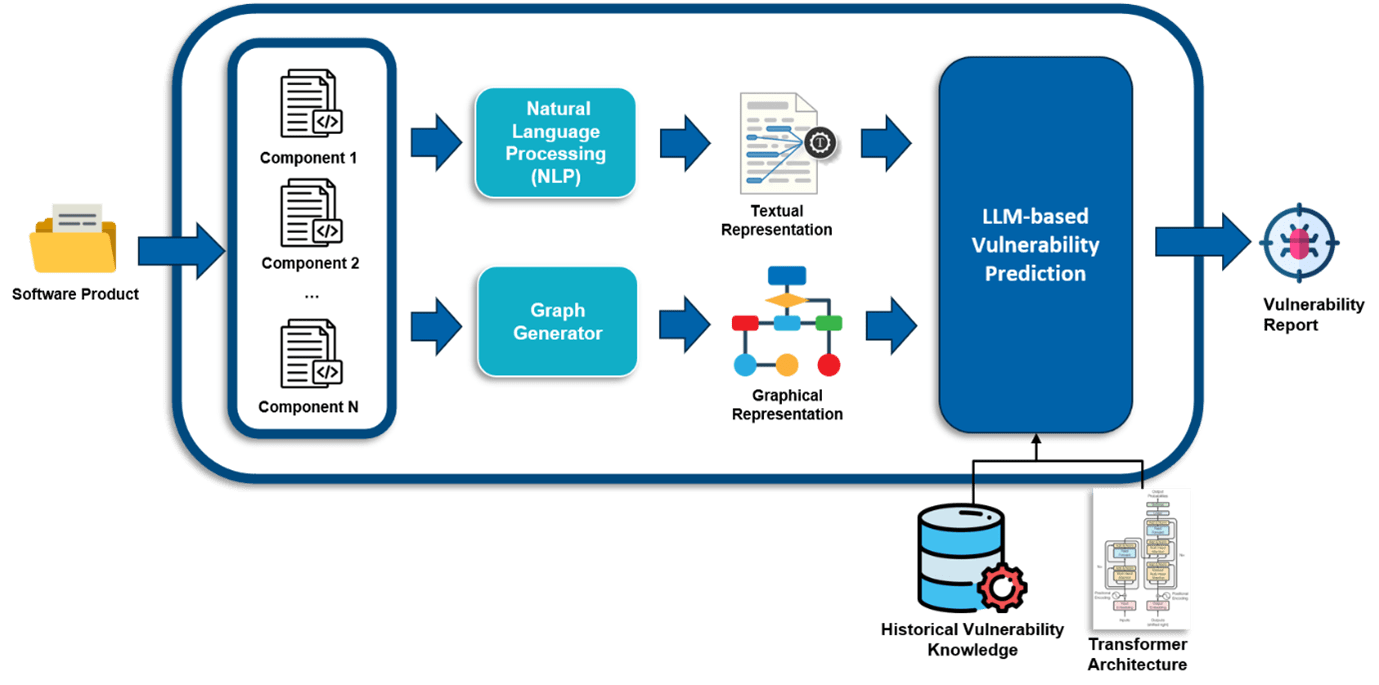}
   \caption{ Description of vulnerability prediction of LLM-based system. Image from \cite{bworld} }

  \label{fig:vulnerability}
\end{figure}
\section{Vulnerability Assessment}
Vulnerability assessment is very crucial component of all the cyber defence strategies. It investigates the whole system for a review of security weaknesses that could be a possible exploit of threat. It provides a systematic help for the  organizations to identify, quantify, and prioritize the vulnerabilities. This includes software, hardware, and network infrastructures to enable organizations to protect the digital assets. The expected outcome of this operation is to alleviate potential risks before cyber criminal exploit them. It involves scanning to the whole system with automated and manual techniques for known vulnerabilities. These vulnerabilities are checked in weak passwords, unpatched software, and insecure system configurations. The results are sort of reports  vulnerabilities' details, such as severity, impact, and recommendations, which will be used to guide security teams to make informed decisions \cite{scarfone2008technical}.

Moreover, vulnerability assessments helps organizations to comply regulations and standards of security controls. This process is recommended in a regular basis to detect new vulnerabilities that may have emerged, as well as confirming the resolution of the previously detected ones. This vulnerability assessments are integral for maintaining the security of organizational environments \cite{eakin2006assessing}. In this section, we review the LLM-based methods of vulnerability assessment. LLMs are used to improve the effectiveness of automatic penetration testing \cite{deng2023pentestgpt} to automate the detection weaknesses in software systems. With the help of LLMs, PentestGPT is able to anticipate security vulnerabilities and propose effective measures to address them. A substantial enhancements in accuracy is achieved compared to conventional penetration testing approaches. This achievement in penetration tests will have significant consequences for the future of cyber security defence measures.
In \cite{ferrag2023securefalcon}, SecureFalcon used LLMs to automate cyber security measures identify, analyse, and counteract cyber threats with speed and accuracy. By incorporating real-time data analysis, SecureFalcon illustrated improved efficacy of security operations along with facilitating ongoing learning and adjustment to emerging threats. Temara \etal investigated the utilisation LLMs, notably ChatGPT, in the early phases of penetration testing to achieve higher performance \cite{temara2023maximizing}. It asserts that LLMs' comprehensive reconnaissance enhanced the results of penetration tests by providing insights about system vulnerabilities. By harnessing LLMs capacities to automate data collection and analysis, penetration testers can achieve the maximum rate of accuracy.

 % \textbf{Vulnerability Detection and Monitoring Using LLM}
Happe \etal used LLMs to find and monitor security vulnerabilities in software systems by leveraging natural language processing skills  and suggest security weaknesses \cite{happe2023getting}. First, LLMs is trained on extensive code datasets, encompassing both secure and insecure samples to identify the vulnerabilities. It showed that LLMs can attain superior accuracy and speed compared to conventional techniques. The use of large language models (LLMs) in penetration testing is examined. It explores the use of LLMs to model attacker behaviour, find security flaws, and evaluate the security posture of apps and systems. Penetration testers can automate a number of testing tasks, such as reconnaissance, exploitation, and reporting, by utilising LLMs' natural language understanding capabilities. The research emphasises how LLMs can improve penetration testing procedures' efficiency that will in return strengthen cyber security defences. The integration of LLMs for zero-shot vulnerability repair is investigated in \cite{pearce2023examining}. Firstly, the concept of zero-shot vulnerability repair is to automatically develop patches for software vulnerabilities without having experienced these vulnerabilities before. It focused at producing patches for software flaws that are known to exist in a variety of programming languages and applications, relying on the ability of LLMs to comprehend vulnerability context and generate syntactically and semantically sound patches. The benefits and drawbacks of zero-shot vulnerability correction with LLMs are considered through testing and assessment, taking into account variables such as patch quality, vulnerability coverage, and possible hazards related to automated patch creation. The detection and monitoring of cyber security vulnerabilities is studied in \cite{akuthota2023vulnerability}. LLMs as machine learning models should be used to analyse textual data, such security advisories, bug reports, and forums to find potential weaknesses in software systems. Various methods are studied to check the ability of LLMs to detect vulnerabilities in the software systems as follows: 1)native language interpretation: utilising the contextual capabilities of LLMs to comprehend and interpret natural language, and hence, recognising the vulnerabilities including error messages, code snippets, and descriptions of program behaviour. 2) pattern recognition: detecting comparable vulnerabilities in new or old software systems by comparing various timely patterns. 3) sensory analysis:  it finds departures from typical behaviour that can point to the existence of a cyber threat. This work emphasises the effectiveness of LLMs  in cyber security monitoring and vulnerability discovery using their natural languages' abilities.

\vspace{2mm}
\noindent
\textbf{Vulnerability assessment of software coding} Ingemann \etal studied the software vulnerabilities using LLMs \cite{ingemann2024software}. According to the results, LLMs have potential advantages over conventional techniques in the areas of functionality evaluation and vulnerability detection, including increased scalability, accuracy, and efficiency. The reason behind this is that LLMs is powered with natural language context. In this work \cite{mathews2024llbezpeky}, LLMs are leveraged to identify vulnerabilities in software coding systems. Since LLMs are adept at deciphering and analysing code patterns and structures, they have shown encouraging outcomes in discovering vulnerabilities. LLMs is integrated in the identification system for vulnerability discovery, particularly to improve the efficacy and precision of security testing procedures in order to strengthen overall security posture. However, This research highlights the necessity of boosting LLMs to encounter obstacles and constraints, including problems like false positives, scaling issues, and the requirement for ongoing training and improvement to meet changing threats. As there is a an increasing concern over security flaws in online applications and in this work \cite{sakaoglu2023kartal}, Sakaoglu \etal presents an novel method, namely Kartal, to identify vulnerabilities such as command injection, cross-site scripting (XSS), and SQL injection in online applications by analysing web application code. The experiments illustrate that the accuracy of Kartal is outperforming the previous methods. Another work of Sakaoglu \etal \cite{sakaoglu2023kartal} to focus on logical vulnerabilities such as SQL injection or cross-site scripting, which result from  poor business logic rather than technical implementation flaws. It utilises refined LLMs to identify patterns suggestive of logical weaknesses in web application code. The proposed method implements an optimised LLMs to identify vulnerabilities by firstly preprocessing data, train on labelled data. Despite the existing obstacles to using LLMs, it significantly improves performance, reduces manual labor, and improves detection accuracy.

In another study of detecting vulnerabilities in smart contracts \cite{chen2023chatgpt},  \etal explored the use of LLMs (ChatGPT) for the purpose of detecting vulnerabilities in smart contracts on blockchain platforms. First, it analyzes the syntax and semantics of contract code to identify any security vulnerabilities. A novel mechanism is provided to optimise using a dataset that have different smart contract codes, as well as documented flaws. It concluded that ChatGPT is able to accurately identify prevalent weaknesses such as reentrancy attacks, integer overflows, and inappropriate access control.

\begin{figure*}[th!] 
\centering 
\includegraphics[width=.9\linewidth]{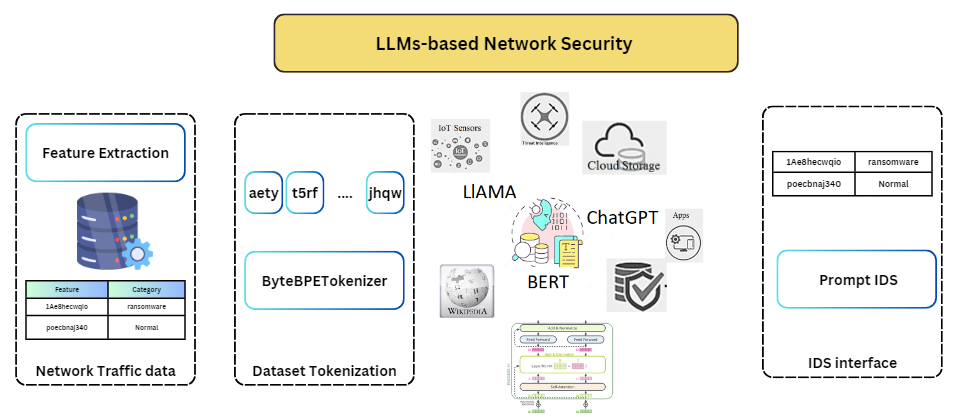}
   \caption{Visual description of the LLMs-based network security steps. LLMs is used to predict the intrusion in the network flow. LLMs is mainly used after fine-tuning.}
  \label{fig:network_security}
\end{figure*}

\section{Network Security}

Network security is a backbone of cyber defence strategies to safeguard network data and infrastructures against unauthorised access, misuse, change, destruction, or unlawful disclosure. In this way, the preservation of the integrity and usability of networking and data resources is guaranteed. It also refers to the range of measures and rules to prevent and monitor unauthorised access, misuse, alteration, or denial of the system's services and its resources \cite{patel2010survey}. 

Intrusion detection is a core element of network security since it is about the identification and response to cyber crimes that target the system or the network. Intrusion Detection Systems (IDS) are primarily used to oversee network traffic and detect abnormal patterns that might cause a network fault or system breach by an intruders. IDS has two main variants: 1) Network-based Intrusion Detection Systems (NIDS), designed to monitor and analyse any suspicious behaviour. 2) Host-based Intrusion Detection Systems (HIDS), implemented for  particular hosts or devices within the network to monitor incoming and outgoing packets and promptly notifies if any abnormal activity, including system calls executed by the operating system or modifying system files. IDS has two main methods: 1) signature-based detection that relies on specific patterns of recognised threats \cite{ioulianou2018signature}, 2) anomaly-based detection, which compares activities to a baseline to identify deviations \cite{jyothsna2011review}. Anomaly-based detection include machine learning techniques. Strong administration is necessary to immediately respond to issues while minimising false positives that might drift the attention from true problems \cite{nedelkoski2019anomaly}. As complexity of cyber threats is getting worse, intrusion detection stands an essential role in network security. Continuous updates and training are necessary to the network traffic data and IDS systems to guarantee effectiveness against cyber threats. Network security is one aspect in the  security plan, as it helps protect critical data from unethical access.

Since the evolution of deep learning, significant improvement has been witnessed in intrusion detection systems (IDS). Deep learning models' power to train massive datasets and learn intricate models boosted IDS to identify abnormalities in network traffic that could escape detection by conventional approaches \cite{rabbani2021review}. It provides the ability to adapt to new threats without the need for explicit retraining from scratch, which help in countering zero-day exploits and advanced persistent threats (APTs) \cite{mirza2014anticipating} \ref{fig:network_security}. DL-based IDS enhances the accuracy of harmful activity identification while diminishing the occurrence of false positives, and hence, the efficiency of cyber defence is significantly improved. In this section, we are reviewing LLM-based network security strategies as we provide discussion on each method, significance, and drawbacks.

\begin{figure}[] 
\centering 
\includegraphics[width=.9\linewidth]{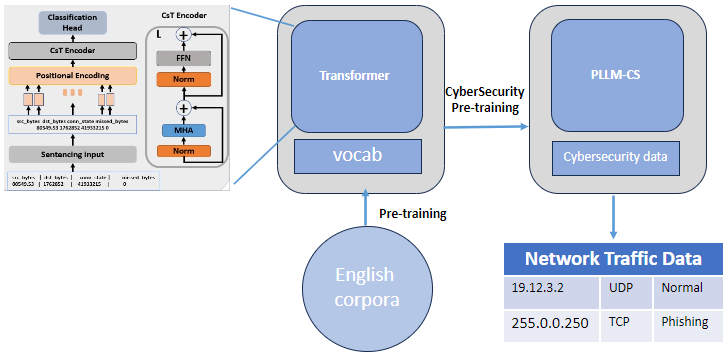}
   \caption{ Fine-tuning BERT model to predict malicious anomalies in satellite networks from \cite{hassanin2024pllm}}
  \label{fig:pllm-cs}
\end{figure}

\vspace{2mm}
\noindent
\textbf{Intrusion Detection} The power of LLMs in recognizing and preventing online fraud is investigated in \cite{jiang2024detecting}. It shows the strength of LLMs in identifying many types of online scams, such as phishing, fake emails, and misleading marketing by training LLMs on large-scale cyber security datasets.
By analysing contextual textual material and using their natural language understanding capabilities through LLMs,  fraudulent behaviour is recognised . The results indicate that LLMs is able  to help detect and stop internet scams because of its ability to classify the infected material from the genuine ones. However, further studies are required to produce robust systems that are able to provide scam identification and mitigation, which in return will ultimately lead to a safer online environment. In this work \cite{shi2024red}, Shi \etal explored the possibility of creating hostile samples with language models that can avoid detection by current intrusion detectors. They show that language models are capable of producing text samples that avoid the detectors while maintaining semantic similarity with the source data. This shows the necessity of constantly improving detection techniques to keep the security safe. It implies that by spotting flaws and providing supervision for developing robust detection systems, merging red-teaming with language models is proved to increase the robustness of LLMs. Hassanin \etal proposed a pre-trained Large Language Model designed for anomaly detection in satellite networks \cite{hassanin2024pllm}. They customized the network traffic data to fit Transformers input, which encode the context in cyber data (see Figure \ref{fig:pllm-cs}). Here \cite{yang2024dlap}, is a similar study but for network traffic data. 
Flowtransformer presents a framework for flow-based network intrusion detection systems (NIDS) that uses transformers as the machine learning model \cite{manocchio2024flowtransformer}. The suggested architecture improves network intrusion detection by capturing contextual data and temporal dependencies inside network flows. The accuracy and efficiency of identifying malicious activity in network traffic is significantly improved with the help of the Transformers \cite{vaswani2017attention}. However, it is still suffers from Transformers issues, including the need of brute force training with huge amount of data as well as this improvement comes on the cost of speed and resources. In this paper \cite{sai2024generative}, the integration of LLMs and cyber security and its impact is investigated in various ways, including threat intelligence, security automation, adversarial defence, and cyber security awareness.
Using LLMs in the landscape of cyber security offers  improvement to the cyber security operations and threat detections. It summarised the benefits and risks of applying LLMs to cyber security operations. In this paper \cite{prasad2023role}, Prasad \etal explores using the pre-trained Transformers, namely ChatGPT, is improving the cyber security tasks. Few areas in cyber security are investigated, including threat intelligence, security automation, adversarial defence, and user education. It draws attention to the model's abilities to analyse massive amounts of text data, automate security chores, identify and neutralise hostile attempts, and provide instructional materials to spread knowledge about cyber security awareness. It also highlights the significance of ChatGPT in improving cyber security principles as well as the ability of mobilization according to the changing threat environment overall. Salloum \etal  explores the techniques used to identify the threats presented by malicious accounts on digital platforms, particularly in the context of generative AI such as ChatGPT \cite{salloum2024detecting}. It uses machine learning algorithms to distinguish between genuine user activities and those that are driven by malicious drives. It highlights the importance of powerful anomaly detection systems to monitor atypical activity patterns and evaluate potential risks in real-time. Moreover, it emphasises the significance of regularly updating systems and educating users to keep up with the ever-changing cyber attacks as it recommends using a multi-layered security strategy to efficiently reduce the risks of fake accounts. 
\begin{figure}[] 
\centering 
\includegraphics[width=.9\linewidth]{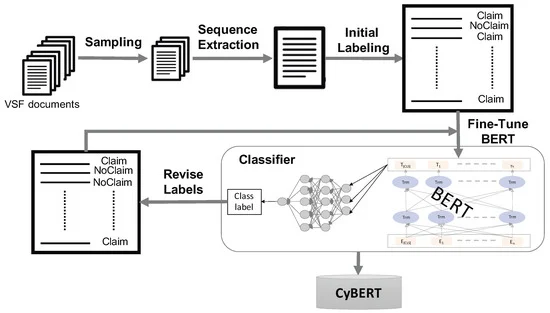}
   \caption{Fine-tuning BERT model to predict intrusions in network data \cite{ameri2021cybert}.}
  \label{fig:cybert}
\end{figure}

In \cite{nguyen2022flow}, Nguyen \etal presented an LLMs-based IDS to analyse network traffic in order to detect anomalies and abnormalities. Because they treated each flow as a word, and a series of flows is a sentence, they enabled the model to accurately predict the behaviour of network traffic. Their work to LLMs is more than a model plugged into their defense system, rather they build a stronger network traffic flows by involving the key elements (source, destination points, protocol flags, packet numbers, etc) into the process. They used real datasets (CIDDS-001 and CIDDS-002 datasets) for model training and validation to consider the upcoming changes in the network settings, therefore, the highlighted the importance language models to tackle the issues of the network flows systems. In this paper \cite{nwafor2022canbert}, Nwafor \etal employed LLMs to detect intrusion detection systems (IDS) in vehicle networks. It builds the whole defensive system around Controller Area Network (CAN) bus systems \cite{hpl2002introduction}. It exploits the susceptibility of these networks to some cyber threats including replay, fuzzing, and DoS attacks to build a stronger model against these networks. Using LLMs in such environments achieved great success because the complexities in the language models. It shown accurate detections for abnormalities and possible risks and superior performance compared to the standard methods. In this work \cite{guastalla2023application}, using LLMs to detect DDoS attacks is investigated. It boosting the machine LLMs as a strong capable machine learning ability to detect DDoS early before spreading its harm to the network. Overall, it is helping the defensive systems to reduce the harm of the DDoS attacks on the network systems. Rieck \etal proposed LLMs model for detection of unknown attacks in the network data \cite{rieck2007language}. It primarily extracts  n-grams and word-based characteristics from network payloads and these data packets and their contents are considered as language units. Then, an linguistic analysis is employed to identify irregularities in the network traffic. One of the noticeable features in this work is that it is unsupervised so it does not require any labelled datasets. In this case, not only this model achieves significant performance on the unlabelled data, it is also capable to detect any abnormality in the ever-changing environments. Also, it achieves such performance in a linear time, which make it very efficient in real-time intrusions detections. Lastly, this work addresses the issue of open vocabulary in the cyber security, which will detect any evasions alterations or obscurity by the attackers in the network traffic. In this work \cite{tuor2018recurrent}, hybrid method of LLMs and RNNs are used to identify the abnormalities in the Vocabulary Event-Level network data logs.  They modeled the network logs on each individual line, utilising the tokens of the character level, This effectively helped to handle the usual diverse languages in such datasets. By modeling a collection of ASCII characters as tokens, it enables the model to depict of any log entry, irrespective of its particular details. Three models are proposed to achieve the goal of abnormalities detection, Event Model, Bidirectional Event Model, and Tiered Event Models, respectively. While traditional approaches face difficulties in handling network data, RNN showed ability to detect cyber threats. Alkhatib \etal proposed an LLM method, namely CAN-BERT, to detect abnormalities in Controller Area Network (CAN) systems, which are frequently employed in automotive networks \cite{alkhatib2022can}. The main objective of this work is to employ the LLM capabilities to analyse CAN network traffic, which is usually non-textual and highly organised. They customized their model to handle continuous data packet streams instead of individual words, dealing with CAN messages as sentences and individual signal values as words. By analysing the CAN messages, it identifies the potential intrusions or malicious activity within the network. The empirical experiments showed that both known and undiscovered attack pathways are accurately detected by CAN-BERT. Li \etal introduces a novel method that combines augmented with pre-trained language models (PrLMs) and Conditional Generative Adversarial Networks (CGANs) to improve intrusion detection systems (IDS) \cite{li2024pre}. The main reason of combining these two methods is to build a sophisticated feature extraction capabilities to improve intrusion detection efficiency. CGANs are used handle the imbalance in the training datasets while LLMs are used to extract complicated pattern of the network features. Overall, this integration of LLMs with CGANs provides a potential way to address inherent issues in network intrusion detection, including the imbalanced datasets with an efficient feature extraction mechanism. It proved that a potential method to significantly enhance the accuracy of detecting abnormalities in the network data.  
Aghaei \etal  present a new method for identifying anomalies such as fraudulent transactions, network intrusions, or atypical user behaviours in data by modifying the Bidirectional Encoder Representations from Transformers (BERT) \cite{aghaei2022securebert}. This variant of BERT improves security and efficiency in detecting outliers in network datasets. Training SecureBERT across various datasets boosted its capability to detect abnormalities. In \cite{piggott2023net}, Piggott \etal presented a chatbot system, namely Net-GPT, that is driven by LLMs and intended to act as a man-in-the-middle (MITM) agent between ground control operators and UAVs. By using LLMs, Net-GPT is able to comprehend and react to operator inputs, offering real-time support and automating certain activities. It emphasises how adding LLMs to UAV systems enhances operational effectiveness and the relationship between humans and machines in UAV. 
Patsakis \etal used LLMs to decipher hidden code that is utilised in actual malware campaigns that is frequently employed strategy by cyber criminals to evade detection by conventional security systems \cite{patsakis2024assessing}. It suggested using LLMs to counteract different obfuscation tactics, including encryption, packing, and polymorphism. It highlights that that LLMs have the potential to simplify complex code structures, however, their usefulness is highly dependent on the complication of the obfuscation methods used as well as the specific characteristics of the virus. It stresses the necessity for additional development of these models to handle such sophisticated obfuscation tactics, as well as the need of continuous training of LLMs to combat with the ever-changing world of cyber threats. In this paper \cite{fujima2023using}, Fujima \etal how to use ChatGPT for ransomware threat analysis and prediction. Ransomware is a kind of malware that causes serious risks to cyber security by encrypting victim's data and demands payment to unlock it. It trains LLMs on datasets containing ransomware messages to recognise the linguistic patterns and traits of ransomware threats. Since ChatGPT is powerful to model the long-range dependencies, it helps recognise common themes, strategies, and signs linked to ransomware attacks. Hence, it showed ability to forecast possible ransomware threats by looking for patterns in past data. Further, it helps cyber security experts to spot new ransomware tendencies and issue cautionary messages. Though the inherited shortcomings of LLMs, it showed significant improvements in deciphering ransomware messages, comprehending threat environments, and forecasting prospective ransomware attacks. 
 % \textbf{Lost at c: A user study on the security implications of large language model code assistants}
Sandoval \etal is identifying security issues that result from using LLMs in code development \cite{sandoval2023lost}. It shows that they are useful in enhancing programming efficiency and precision, however, they might unintentionally introduce security vulnerabilities in code generation. Pointing out security issues of LLMs in code development helps the personnel to avoide it as well as creating the awareness amongst the employees.

 % \textbf{Using LLMs for secure code development without compromising functionality}

In \cite{he2023large}, He \etal explores the capacity of LLMs to assist in software code evaluations, with a specific emphasis on identifying security vulnerabilities as well as evaluating code functionality. Firstly, this work investigates how LLMs can optimise the code review process, saves a lot of software development costs. They found that by automating certain aspects, LLMs have the ability to enhance efficiency and minimise human error. LLMs helps in vulnerability identification as a machine learning module. Though it showed the significance of LLMs in protecting, reviewing and identifying vulnerabilities for software programs, it need a training customization on this paradigm.  In \cite{okey2023investigating}, Okey \etal investigates LLMs and cyber security on topic modeling and sentiment analysis. In order to provide insights into new threats and vulnerabilities, topic modelling seeks to discover popular cyber security themes and patterns within online discussions. Sentiment analysis evaluates the general attitude expressed in discussions on cyber security, which aids in determining public opinion and identifying possible warning signs of danger. The study illustrates ChatGPT's potential to improve cyber security intelligence and strategies by utilising its natural language processing capabilities.  In \cite{sharma2023impact}, Sharma \etal  explores the combined influence of big data analytics and LLMs on cyber security practices. It delves into how these technologies enhance threat detection, incident response, and the other security operations. Since big data enables organizations to process vast amounts of data, it helps identifying anomalies and potential threats. The main reason behind this improvement is that LLMs enhances natural language understanding and facilitates automated responses in security operations. Hence, it concludes that LLMs empower organizations to strengthen their cyber security posture, better response incidence, and effective  threats detection. In \cite{lai2023intrusion}, Lai \etal uses the natural language processing features, LLMs, counteract the increasing spread and sophistication of cyber threats. It uses LLM to analyse the network data and identify any abnormality in the patterns and due to the power of LLMs in pattern recognition, it achieves high performance in the field of intrusion detection. Overall, this work presents a crucial method for large-scale language model intrusion detection and hence it improves the accuracy, scalability, and flexibility in response to changing cyber attacks.

\noindent
\vspace{1mm}
\textbf{Email Phishing}
Labonne \etal proposed using T5 (Text-to-Text Transfer Transformer) in identifying email spam through few-shot learning methods \cite{labonne2023spam}. A performance comparison between T5 and the other LLMs in detecting spam emails when there is a scarcity of labelled training cases is provided. The results illustrate the superiority of T5 over the other LLMs in detecting spams with less labelled data, as well as exhibiting resilience when dealing with different types of spam. Roy \etal discussed the increasing concern about the use of LLMs such as ChatGPT, Google Bard, and Claude in the creation of complicated phishing scams that utilise AI-driven chatbots to generate challenging phishing attempts \cite{roy2023chatbots}. The main purpose is to learn the cyber attacks ways to reduce their risks, such as creating AI detection tools capable of identifying phishing attempts, as well as enforcing stricter regulations on the use of generative AI technologies. In \cite{wu2024evaluating}, Wu \etal compare the performance of ChatGPT and conventional spam detection algorithms, more precisely the  capacity to handle the subtleties of natural language in spam text. ChatGPT surpassed traditional approaches in accuracy and speed, due to its sophisticated natural language capabilities. Further, spam techniques that frequently evaded traditional filters have been detected by ChatGPT. It also reduces the occurrence of false positives that recommend it to be plugged in email systems. Heiding \etal investigated the proficiency of LLMs in producing and identifying phishing emails compared to human \cite{heiding2023devising}. It showed that LLMs in generating and detecting phishing emails are more efficient than humans due to their comprehension of linguistic subtleties. However, LLMs are identified as double faces for cyber security, it can be defensive assets or dangerous vulnerable sources. Li \etal combined LLMs with multi-modal knowledge graphs to improve the accuracy of reference-based phishing detection of phishing attempts \cite{li2024knowphish}. By utilising textual and non-textual components (such as graphics and links) , KnowPhish  authenticates information using a comprehensive knowledge network that encompasses verified data on authentic websites and visual identities. Unlike conventional methods that identify phishing based on one technique, KnowPhish uses system's architecture that integrates the predictive capabilities of LLMs for text analysis knowledge graphs for evaluating the credibility of emails and web pages. The performance of KnowPhish is proved to be better in identifying complex, context-based phishing schemes that use many deceptive strategies. In \cite{hur4815382utilizing}, LLMs are employed to identify SMS spam through a few-shot learning methodology; that is, using a limited number of instances to tackle the difficulty of quickly adjusting to new and changing spam strategies without retraining.  Pre-training LLMs showed the ability to generalise from a limited number of instances and reliably categorise SMS messages as either spam or non spam. This shows  the efficiency of LLMs on few-shot scenarios compared to conventional machine learning techniques that often necessitate bigger training datasets. Such performance indicates that LLMs not only improvise without annotated datasets but also higher performance in identifying spam. This approach has substantial ramifications for the security of mobile communication, providing a salable and flexible solution to efficiently tackle spam messages sent over SMS. Koide \etal used LLMs to recognise phishing emails by transforming emails into prompts to check them \cite{koide2024chatspamdetector}. This makes it possible to look more closely at the content of the email in its context and spot sophisticated social engineering techniques. Because it can look for differences between the brand name and the domain name associated with it, it spots brand impersonations effectively. They introduced a dataset of phishing emails collected from honeypots between August 2022 and October 2023. These emails have no URL in the body, which effectively removed any emails that attempted to send recipients to phishing websites. It targeted 193 brands in 19 languages—English, Portuguese, German, and Dutch. Though it is promising for identifying phishing emails, it is crucial to take into account the constraints of LLMs and the computing expenses. Patel \etal investigated using LLMs in identifying phishing emails, and they performed a comparative analysis between LLMs and conventional machine learning-based phishing detection systems \cite{patel2024large}. They highlighted that LLMs has advanced learning capacities that excel detecting phishing emails, especially after training on various datasets. The results indicate that LLMs not only achieve better accuracy, but also provide resilience against new phishing techniques. Trad \etal provided a thorough comparison between two approaches, prompt engineering and fine-tuning, in utilising LLMs for detecting phishing attempts \cite{trad2024prompt}. It is an important study because the advancement of LLMs is turning the machine learning as a blackbox.  First, LLMs are given specially designed prompts or tested after fine tuning of phishing and lawful messages' training. It highlights that prompt engineering is computationally efficient and faster, however fine-tuning provides superior accuracy, as well as flexibility to new phishing strategies. Therefore, the recommendation is to use them according to the case or the problem. Nahmias \etal proposed LLM-based method to improve the detection of spear-phishing attacks by utilising prompted contextual vectors \cite{nahmias2024prompted}. Because Spear-phishing is an advanced way of stealing sensitive information by sending false emails that imitates a trustworthy sender,  detecting spear-phishing is very sophisticated task. The main contribution of this work is that it utilises the contextual embeddings by language models. These fine-tuned contexts are produced using specialised prompts to assist in identifying spear-phishing. This is the first study that incorporates prompted contextual vectors into current spear-phishing detection algorithms. They succeeded to differentiate between authentic messages and malicious ones by training on a large dataset. Another study \cite{jamal2023improved} that presented an advanced algorithm to improve the performance of identifying hazardous emails, such as phishing attempts, spam, and genuine (ham) communications. It is mainly providing a strong training on large datasets that uses a wide range of real-life email conversations to boost the accuracy efficiency in different situations. In \cite{heiding2024devising}, Heiding \etal explored multifaceted use of large language models (LLMs) in both creating and identifying phishing emails. They generated phishing emails by LLMs that look authentic as well as they used the same way to detect phishing emails by training on large datasets. Yu \etal used LLMs to improve security measures through the creation of a concept called honeywords which provides a combination of false passwords and the user's real password to baffle attackers as well as identify security breaches when they try to utilise these wrong passwords \cite{yu2023honey}. Therefore, the generative capability of LLMs is used to generate these honeywords, which complicate the task of attacker to know the authentic words. Despite the effectiveness of the honeywords concept, they rely on the training data and the model's capacity to generate false yet plausible alternatives. In \cite{chataut2024can}, Chataut \etal uses LLMs in spotting phishing attempts. Phishing attacks are serious cyber security risks that trick people into disclosing personal information via using social engineering techniques. The ability of LLMs to use contextual signals and language patterns to analyse email content with the aim of detecting phishing attempts is investigated. It uses LLM to boost the ability of their model in differentiating between phishing and legitimate emails by training them on a dataset of those emails. The results show the efficiency of LLMs to detect email phishing and secure emails. Overall, LLMs illustrate significant improvment regarding email phishing. An extensive study on log anomaly detection using Transformer-based Large Language Models (LLMs) is explored in \cite{balasubramanian2023transformer}. Firstly, Transformer-based LLMs is used mainly to identify irregularities in log data by detecting unusual patterns that could be signs of system failures. Also, LLMs-based chatbots are used to facilitate verbal engagements between potential attackers and malevolent actors to acquire information that divert the attention of attackers. This study showed that LLMs is capable of overcoming the cyber attacks and boost efficiency. In this paper \cite{ferrag2024revolutionizing}, Ferrag \etal proposes a novel approach to improve cyber threat detection in IoT (Intenet of Things) and IIoT (Industrial Internet of Things) systems. It presents LLM-based privacy-preserving method for sifting through device communications and data streams to look identify possible cyber attacks. With limited resources, the proposed method proved significant performance in anomaly detection, security monitoring, and intrusion detection in IoT/IIoT. Overall, it proposes an effective approach to strengthen cyber security in IoT while taking resource constraints and privacy issues into account. Ameri \etal proposed using a refined iteration of the BERT model for detection of abnormalities present in text \cite{ameri2021cybert}. It refined it by training on the specific datasets of cyber security such as  as shown in Figure \ref{fig:cybert}. It showed superior accuracy results compared to the conventional machine learning methods. Bayer \etal provided a tailored BERT model for the cyber security field to  analyse the specific terminologies in cyber security writings, encompassing technical threat reports and security protocols \cite{bayer2024cysecbert}. Ranade \etal proposed CyBERT to improve the identification cyber security-related issues in text, including various types of malware, attack methods, and security weaknesses across reports, logs, and research articles \cite{ranade2021cybert}. In order to fit it to cyber security, they trained BERT on the specialised datasets, and thus, improving the performance of abnormalities detection. In \cite{wohlbach2024evaluating}, Wohlbach \etal evaluates the possible cyber security threats of NLP models, including ChatGPT and Google Bard. It discusses the ways that bad actors may use these models to commit phishing scams, disinformation campaigns, and data breaches, etc. The aim is to increase public knowledge of the possible cyber security issues of using NLP models in real-world applications by exposing their vulnerabilities and shortcomings. Furthermore, it offers valuable perspectives on approaches to mitigate cyber security risks of LLMs models, including model validation, adversarial learning, and self-security focused model design. In \cite{zaboli2023chatgpt}, Zaboli \etal investigates improving cyber security measures using  large language models (LLMs), particularly chatGPT for smart grid applications. It looks on how LLMs  capabilities can be used to detect and mitigate cyber security risks in smart grid networks. It demonstrates using LLMs to evaluate and decipher intricate data patterns, find abnormalities, and offer practical insights, and thus, improve smart grid security. It provides opportunities, difficulties of  integrating LLMs into smart grid cyber security plans, aiming to infrastructure resilience against cyber threats.

\vspace{2mm}
\noindent
\textbf{LLMs-based Honeypots} Mckee \etal explored the use of LLMs to improve honeypot technology in order to counter command-line cyber-attacks \cite{mckee2023chatbots}. Koide \etal explored the utilising LLMs to detect phishing websites to identify deceitful language and misleading content commonly seen on phishing websites \cite{koide2023detecting}. It analyses linguistic patterns and irregularities in online website text as it asserts that incorporating LLMs into cyber security frameworks boost significantly the identification of phishing activities. Sladi \etal investigates using LLMs's generative power to generate dynamic, persuasive, and interactive environments that imitate authentic systems in order to deceive and engage attackers \cite{sladic2023llm}. Thereby gaining valuable knowledge about attack techniques, which help in detecting the cyber attacks. Its primary way is to produce authentic network traffic, user interactions, and system logs in real-time.  These LLM-based honeypots are more efficient in attracting attackers for extended durations, hence offering more profound insights into malevolent approaches.

% % \section{Incident Response and Management}

\begin{figure*}[] 
\centering 
\includegraphics[width=.9\linewidth]{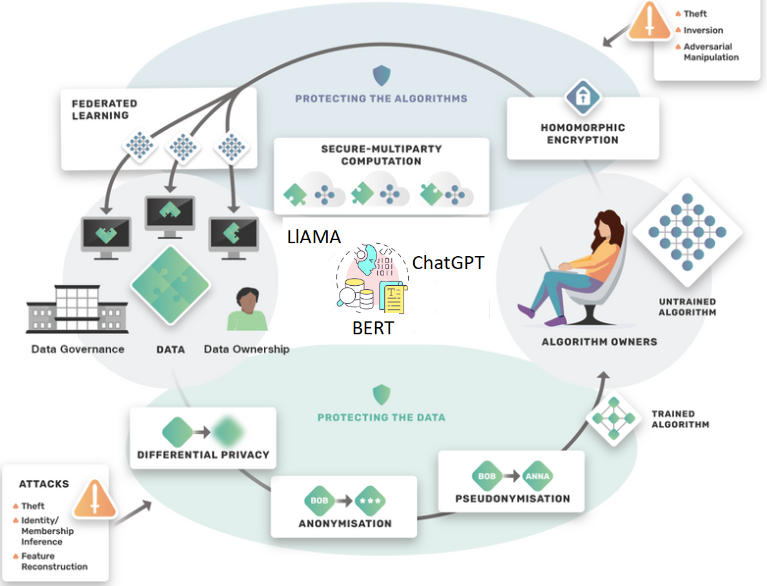}
   \caption{Visual description of the LLMs-based privacy preservation techniques. LLMs can be integrated in all the systems of privacy preservation such as data privacy, and algorithm protection.}
  \label{fig:pp_llms}
\end{figure*}

\section{Privacy Preservation}
Preserving privacy is an integral  part in protection of personal and organisational data from unauthorised access. Lately, it became crucial for both individuals and companies to protect sensitive information, confidentiality and integrity, particularly after the progression of AI models and the vastness of digital systems. However, it requires simultaneously to implement strong security measures to counter potential cyber threats. Diverse range of approaches precisely designed to safeguard personal data and guarantee exclusive access by authorised users. For instance, encryption is a fundamental technology in this field, employed to encode data in a manner to restrict access to only those individuals with the appropriate decryption key. Also, anonymization and data masking techniques are used to eliminate personally identifying information from data collections, thus alleviating data breaches' issues.

With the evolution of deep learning models, although their impressive analytical powers, problems regarding the protection of privacy due to their reliance on extensive datasets have risen. In order to tackle these concerns, advanced methods like differential privacy, federated learning, and homomorphic encryption are being more and more included into deep learning frameworks \cite{wang2021deep}. Stochastic perturbation to obfuscate individual data points have been used in \cite{kargupta2005random}. Federated learning have been introduced to enable the training of models on distributed devices, which protects confidentiality of data by training in its original place. These strategies aid in reducing privacy issues, enabling the use of deep learning's advantages while ensuring the protection of user privacy. However, more concerns have been introduced due to the evolution of LLMs that provide more challenges to the cyber defence privacy preservation \cite{chen2023can}. Privacy preserving still needs a lot of attention as many gaps need more investigation as Figure \ref{fig:pp_llms}. In this section, we review the recent works on privacy preservation and LLMs.

Kim \etal used LLMs to address a significant concerns about the leakage of personally identifiable information (PII) while training on web-collected data, which include sensitive personal information \cite{kim2024propile}. It suggests that data subjects to test LLMs with a specific information to check whether their information will be leaked. This work handles "black-box probing" where they can check if indviduals' PII is likely to be leaked, and  "white-box probing" to check if provider of LLM to improve the model’s handling of private information. In \cite{raeini2023privacy}, Raeini \etal provides a unique approach to large language models (LLMs) called privacy-preserving large language models (PPLLMs) to protect user privacy without compromising the efficiency. To guarantee that sensitive data stays private and secure during training and inference, these models used federated learning, differential privacy, and secure multi-party computation. It solves privacy preserving concerns including data privacy, confidentiality, and security by integrating privacy-preserving techniques with LLMs. One of the main issues of LLMs is the adversarial attacks and lack of security, therefore, PPLLM  is to find a middle ground between protecting user privacy and utilising big language models for a variety of purposes, such as conversational bots, text creation, and cyber security. Though it shows improvement in the privacy preserving, it comes on the account of accuracy due to the encryption layer. Another explanatory study \cite{gupta2023chatgpt}, generative AI models' impact on cyber security and privacy preserving is investigated. In threat detection and Intelligence, generative AI models are used to examine big datasets and find patterns suggestive of cyber threats, and as a result, threat detection capabilities can be improved. They also delved into the possibility of adversarial assaults directed at generative AI models. In privacy problems, they highlighted that generative AI models gives rise to privacy problems because there is a chance that they will unintentionally produce private or sensitive data, endangering people's privacy. For this, mitigation methods, such as applying privacy-preserving approaches and adding protections against adversarial attacks, are suggested to address these vulnerabilities. Exploring the ability of LLMs to dispel common misconceptions in order to provide advice on security and privacy-related issues in \cite{chen2023can}.  LLMs' capacity to respond to enquiries about security and privacy, including falsehoods like those regarding password security or data privacy procedures is investigated. The results show insights into the efficiency of LLMs as instructional resources in the fields of cyber security and privacy. However, it highlights the issues of this model that draws the attention of the research community to conduct additional research and make improvements in order to increase LLMs' capacity to a precise and trustworthy advice on security and privacy-related issues.  In this paper \cite{singh2024enhancing}, Singh \etal studied the range of privacy and security issues and proposed LLMs to address them. The proposed method is simply to use zero-knowledge proofs without disclosing any particular information. They validated the integrity and accuracy of the model without revealing the underlying data or model parameters.

\section{Awareness}
Awareness of the organisation staff is a crucial component of the important cyber defense techniques, strengthening the need of educating individuals and organisations about the dangers of cyber threats and how to avert them. The main objective of this is to establish a security-oriented environment where all individuals comprehend their responsibilities and roles in protecting digital resources against cyber risks \cite{galinec2020design}. Awareness programs are essential due to human mistakes continue to be a prevalent vulnerability exploited by cyber criminals. Such programs usually include several topics, including fundamental of cyber security, different types of threats, ranging from phishing, to ransomware, secure practices, personal secure networks tips. It also includes training which is frequently covering sophisticated subjects such as mobile security, and the secure use of private social media \cite{aleroud2017phishing}. Comprehensive awareness programs not only provide staff with required knowledge to avoid threats, but also security procedures that can minimize the risk of a cyber crimes. However, consistent upgrades and  courses are essential to keep up-to-date with the ever-changing cyber threats \cite{lagana2018information}. In this section, we are reviewing LLM-based network security strategies as we provide discussion on each method, significance and drawbacks.  \cite{gundu2023chatbots} offers a strategy for using ChatGPT \cite{roumeliotis2023chatgpt} to improve information security behaviours. It identifies the issue of being individuals frequently become targets of cyber attacks because they do not pay attention to security best practices. It addresses this issue through ChatGPT's features to deliver security reminders and nudges. For instance, it remind users to reinforce secure practices, such as creating strong passwords and using two-factor authentication; provide personalised security advice. While this framework seeks to foster an awareness of security within organisations, it is a theoretical approach and lacking the evaluations. In \cite{yamin2024applications}, Yamin \etal  utilised LLMs to automate the generation of authentic and intricate scenarios for cyber security training exercises that will be used for training modules to be more robust against cyber threats. These cyber security incidents and exercises can be used to train the cyber security teams to acquire knowledge and produce  plausible training scenarios. It recommends incorporating LLMs into cyber security training programs that will offer ongoing learning and adjustment to emerging cyber threats.
\begin{figure}[th!] 
\centering 
\includegraphics[width=.9\linewidth]{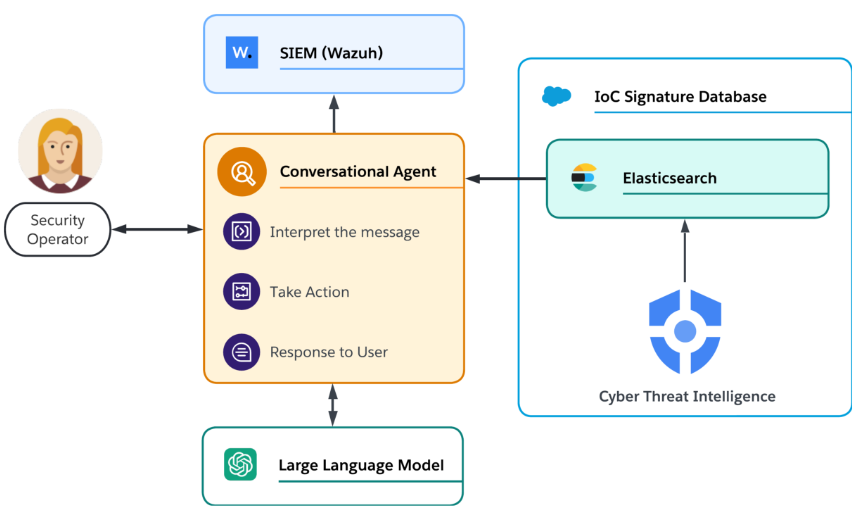}
   \caption{ The proposed LLM-based method for developing awareness of security issues \cite{kaheh2023cyber}.}
  \label{fig:cyber_sentinel}
\end{figure}

This work \cite{kaheh2023cyber} looks into the usefulness of GPT-4 \cite{kalyan2023survey} powered conversational agents in streamlining security-related activities. It investigates how conversational agents can help with different security activities based on superior language capabilities. It proved that it leads to the simplification of procedures like threat analysis, incident response, and security policy enforcement. The study emphasises how cyber security workflows can benefit from the use of LLMs agents to improve productivity and efficacy in security operations management (see Figure \ref{fig:cyber_sentinel}). The impact of large language model (LLM) chatbots in boosting cyber threat awareness through Open Source Intelligence (OSINT) is investigated in \cite{shafee2024evaluation}. It evaluates the abilities of LLM-based chatbots to decipher and analyse OSINT data to deliver threat detection in time, spot possible cyber security threats, and instruct users on security incident response. It is mainly studies how LLM chatbots perform in providing users with precise and useful insights that in return will raising the level of awareness regarding cyber threats and intrusions.

\section{cyber security operations automation}
Automating cyber security operations has become an essential requirement in defending the recent digital systems, particularly with the constant changing in cyber threats \cite{mohammad2018security,kinyua2021ai}. with AI and LLMs, security applications can improve their capacity to efficiently identify, address, and minimise security problems without human intervention. Such operations utilise advanced technologies and platforms capable of executing duties often carried out by human analysts, such as identifying threats, responding to incidents. This automation not only speeds up the process of resolving threats, but also decreases the likelihood of human mistakes \cite{sarker2024ai}. Additionally, humans can not handle massive amounts of data as they get boring quickly, however, automated cyber operations can address this issue and help fix the issue before growing up.

Implementing automation in cyber security operations also tackles the increasing problem of the cyber security skills shortage and lack of experience. Skills shortage can cause harm to the organisation due to the late response, which may cost massive loss. Automation after the last progression of LLMs, is a doable task, especially with repetitive duties, which in return will free up the security staff to concentrate with complicated tasks \cite{mohammad2018security}.

Security Orchestration, Automation, and Response (SOAR) technology offer a smooth integration of different security tools on automated workflows, making it easier to automate responses to incidents \cite{bartwal2022security}. This helps enhancing the efficiency of security operations, particularly in the absence of skills. The last progression of AI and GPT models open the wide door for cyber security operations to rise up and close the inherent gaps. In this section, we will provide overview for the crossed steps of LLMs in the automation of cyber security operations.

Hays \etal investigates the application of LLMs, ChatGPT-3, in improving incident response methods to automate the crucial procedures of documentation, analysis, and review \cite{hays2024employing}. It uses historical data and predictive analytics of LLMs to produce incident reports, delivering immediate decision assistance, and proposing measures to address breaches. The main benefits LLMs is to enhance responsiveness and precision, together with the capacity to overcome emerging risks and scenarios. It also tackles obstacles such as guaranteeing the dependability of advise given by AI and the susceptibility of AI systems to manipulation by malicious individuals. It recommends guidelines to incorporate LLMs into current security operations centres (SOCs) and emphasises, particularly with continuous training to optimise the efficiency of these AI tools in incident response against ever-changing cyber attacks. In a study to use LLMs to defend the web sites from automatic hacking by executing automated cyber-attacks on them \cite{fang2024llm}. First, they utilise the sophisticated natural language processing and machine learning methods to identify weaknesses in online infrastructure, such as SQL injections and cross-site scripting problems without the need for human involvement. Then, cyber defence mechanisms will be used to predict and reduce AI-powered threats, thus protecting digital assets from automated advanced attacks.

In \cite{feffer2024red}, Feffer \etal explored the creation of a sophisticated LLM tailored for the purpose of red teaming in order to assess and enhance its defensive capabilities. The main purpose is to automate and the cyber security operations, hence improving their resilience against a wide range of cyber threats. The results illustrated the model's capacity to detect vulnerabilities with greater efficiency. It also discusses potential security issues associated with automating red team attacks that will highlight the significance of automation of cyber security operations. plugging-in LLMs into an automated IDS is discussed in \cite{g2024harnessing}. First, network data are utilised to interpret large quantities of network data and records to identify abnormal patterns. The main feature of this technique is that it promptly respond to emerging threats. Moreover, it demonstrated the superiority of the LLM-based IDS over traditional systems in identifying evolving cyber threats. Sultana \etal explored the use LLMs for automating cyber operations in different scenarios, including threat identification, system surveillance, and incident handling \cite{sultana2023towards}. It emphasises the ability of LLMs to improve the effectiveness of LLMs in cyber operations to reduce the risks. They proposed merging LSTM with LLMs to combat the real time and ever changing cyber threats. They highlighted  the importance of involving cyber security experts in the process of LLMs to guarantee their conformity with security norms and prevent the introduction of additional vulnerabilities.
One-day software vulnerabilities  that are revealed but not yet fixed have been investigated \cite{fang2024llm} for detection using LLMs. Fang \etal trained LLMs on a large collection of security patches, vulnerability reports, and exploit code to examine publicly published weaknesses and developing appropriate exploit code. The experiments indicate that LLMs, when provided with comprehensive vulnerability knowledge, are capable of mitigate operational risks of attack scripts.  Using LLMs in detection of hardware trojans that cause alterations to hardware circuits is proposed in \cite{kokolakis2024harnessing}. It shows the ability of LLMs to not only improving security protocols by predicting trojan assaults in hardware, but also identifying flaws within these systems. Notable enhancement in the ability to detect hardware trojans at an early stage is achieved by LLMs compared to normal methods. This work is very important for safeguarding the hardware components as it illustrates the significance of LLMs-based cyber threats' detections. Helbling \etal proposed using LLMs to as a self-examination mechanism to analyse their own responses and the the prompts \cite{helbling2023llm}. The methodology combines heuristic-based and machine learning techniques to enable LLMs to evaluate their operational integrity in real-time. The results showed that this self-examination of LLMs boosted the model's ability to counteract adversarial attacks. It suggests that self-aware LLMs with explainable characteristics are necessary for AI ethics and governance to guarantee interpretability. 
Attack trees are hierarchical structures that are used to depict possible attack scenarios and system weaknesses. In \cite{gadyatskaya2023chatgpt}, Gadyatskaya \etal uses ChatGPT's to automatically create attack trees in order to analyse textual descriptions of cyber security risks. By streamlining the attack tree generation process, this method hopes to give cyber security experts an invaluable tool for threat analysis and threat mitigation. As a result it shows how LLMs used to automate difficult cyber security jobs as well as they are employed to boost the effectiveness of threat detection and response incidence. Shao \etal used LLMs to tackle one of the main offensive security tasks, Capture The Flag (CTF) challenges that target computer security issues \cite{shao2024empirical}. Two distinct procedures have been investigated for solving CTF challenges utilising LLMs. The first utilises a human-in-the-loop (HITL) approach, while the other is completely automated, which  explains the effectiveness of LLMs into the problem-solving process. They showed that these models can perform at least as well as, if not better than, human participants in specific situations. David \etal discussed automating the smart contract security by using LLMs, particularly in the presence blockchain technology \cite{david2023you}. This work investigated the efficacy of LLMs in identifying vulnerabilities and compares them to conventional manual auditing techniques. They concluded that although automated technologies have made great advancements and can detect numerous common vulnerabilities, they are not yet capable of completely replacing the nuanced comprehension human experts. The authors assert that manual audits continue to be a crucial element of smart contract security, especially for  guaranteeing adherence to legal norms. Cambiaso \etal  used LLMs to automate responses to fraudulent emails, aiming at assessing the ability of LLMs to interact with scammers and discourage their activities by generating contextually suitable replies \cite{cambiaso2023scamming}. By imitating ordinary human reactions, LLMs waste the scammers' time, resources, and diminishing the effectiveness. They highlighted the necessity of human supervision to prevent unforeseen outcomes, such as the continuation of detrimental content, especially after the scammers recognise they are automated. This study indicated that valuable new instrument in combating email-based scams is provided by LLMs.

\section{Ethical LLMs}

\begin{figure*}[th!] 
\centering 
\includegraphics[width=.9\linewidth]{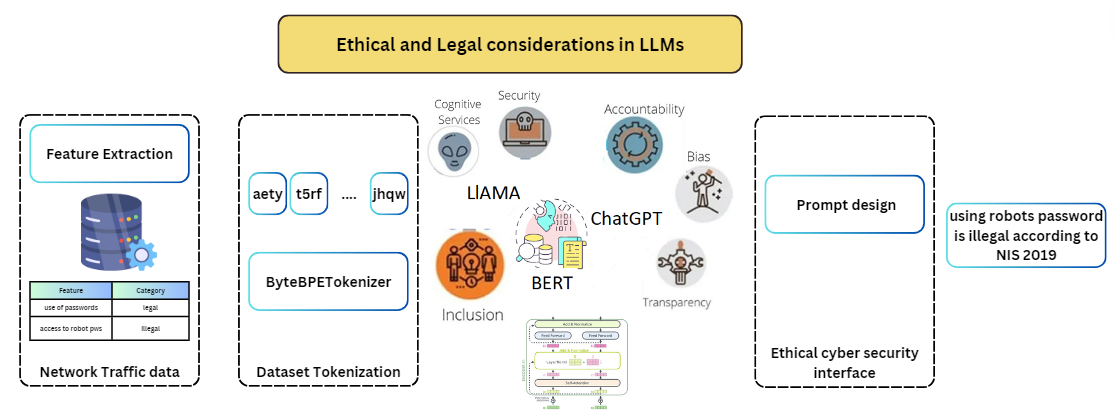}
   \caption{Visual description of the LLMs-based ethical and legal frameworks.}
  \label{fig:ethical}
\end{figure*}

Ethical cyber defense is very crucial part in the cyber defense process in the organisation. It involves the implementation of guidelines, concepts and procedures to protect digital assets and information systems from any cyber threats. With the significant spread of artificial intelligence and its availability to cyber criminals,  the focus on the ethical aspects of how these technologies have to be managed. Ethical cyber defence signifies the value of protecting user privacy, fairness, and transparency in the creation and implementation of cyber security measures. A comprehensive structure that surpasses basic adherence to laws and regulations is entailed under ethical cyber defense, as well as a dedication to making ethical decisions at all levels, ranging from technology selection to to respecting ethical limits. For instance, it may be possible to oversee all employee actions on a company network to avert data breaches, however, ethical factors should dictate the manner of such monitoring. Simply put, ethical cyber defence is the proactive obligation of organisations to implement the cyber defense process along with ensuring that no harm to others. Therefore, it encompasses the responsibility of revealing vulnerabilities, exchanging threat intelligence, and refraining from selling software with a possible exploitation. To provide machine learning-based cyber defense, the fairness of AI decisions have to be fair and just. One of the ways to provide that is to use explainable AI (XAI) that is responsible for justifying the AI conclusions and decisions. In this section, we are summarising the methods of cyber defense with large language models in the recent literature. Figure \ref{fig:ethical} shows the common design of ethical LLMs for cyber security considerations.

\begin{figure}[th!] 
\centering 
\includegraphics[width=.9\linewidth]{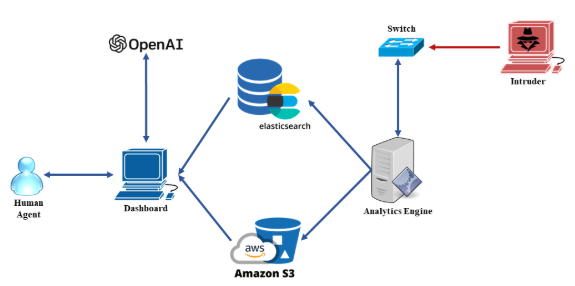}
   \caption{ The proposed LLM method to provide ethical frameworks for cyber security teams \cite{ali2023huntgpt}.}
  \label{fig:huntgpt}
\end{figure}

In a explainable LLMs study, Ali \etal  presents a model that combines a Random Forest classifier, LLM in conjunction with XAI tools such as SHAP and Lime to detect anomalies in a high performance manner \cite{ali2023huntgpt}. LLMs to detect abnormal patterns or anomalies in network traffic, XAI frameworks to comprehend the outcomes of the LLMs (see Figure \ref{fig:huntgpt}). In \cite{uddin2024explainable}, Uddin \etal introduced pre-trained Transformers to identify phishing emails, as well as explanations for its judgements, thereby improving the transparency and reliability. The different part here from previous methods is the incorporation of explainability mechanisms, which enable users to comprehend the underlying reasoning behind classifications. It shows that the model outperforms traditional machine learning methods in terms of accuracy and explainability.

\begin{figure}[th!] 
\centering 
\includegraphics[width=.9\linewidth]{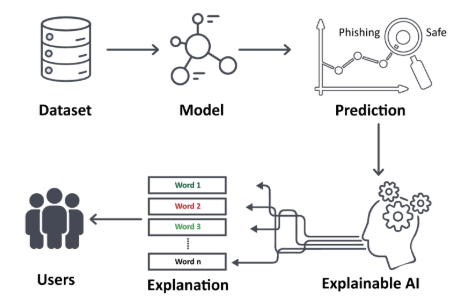}
   \caption{ The proposed LLM method of transparency and explainability \cite{uddin2024explainable}}
  \label{fig:xai}
\end{figure}

In an exploratory study \cite{sebastian2023chatgpt}, Sebastian \etal investigates a number of cyber security areas, including data privacy, adversarial assault susceptibility, the possibility of disseminating false information, and social engineering strategies in the presence generative AI. It highlights the need for thorough risk assessments and security procedures to reduce potential risks. Also, it sheds light on the security issues associated with the deployment of AI chatbots. It proposes creating a more secure environment for the deployment of AI chatbots by bringing attention to handle the cyber security threats.
Garvey \etal explored the potential use of LLMs, mainly ChatGPT and Bard, as avatars for red teaming in foresight scenarios \cite{garvey2023can}. It uses LLMs to automate the process of using an adversarial method to assess plans and tactics to check how far LLMs to replace humans. It presents a comprehensive overview of the advantages and constraints of employing LLMs in this particular context. It proved that LLMs has the ability to significantly improve the complexity of scenarios,however, under the supervision of humans.

% \end{itemize}

\section{Challenges and Open Problems}
Despite the noticeable success of LLMs beyond scientific research and applied sciences, they still suffer from several challenges that need to be fixed to obtain the maximum benefit from LLMs. The main impediments include susceptibility to bias, as LLMs can perpetuate existing biases present in training datasets, as well as privacy leakage concerns. Additionally, high computational resources along with massive amounts of data are needed to train from scratch or fine-tuning. Although these models can be customized to downstream tasks, their performance is not as efficient as training from scratch. There have also been some challenges in explaining and interpreting LLMs' decisions and learning. In this section, we provide an overview of these challenges and limitations, recent efforts to tackle these limitations, and discussing the open research gaps.

\vspace{2mm} 
\noindent   
\textbf{Customization:}  First, because LLMs are trained on general-purpose broad datasets, cyber security context is very limited, which limit their effectiveness in detection and respond to cyber threats. Fine-tuning techniques are proposed in the literature to address this issue, however, it is still challenging due to the scarcity and sensitivity of high-quality labeled data \cite{kumar2022fine}. Additionally, making sure that these customization provide integrity to whole task is another concern. The last noticed attention from research community to this point probably provide promising gate towards fixing this issue.

\vspace{2mm} 
\noindent   
\textbf{Cyber attacks' ever-changing landscape:} 
Cyber threats witnesses continuous change in complexity and frequency with rapid development of advanced attacks including as phishing, ransomware, and advanced persistent threats (APT) \cite{dillon2021cyber}. Such a landscape limits the ability of LLMs to provide ultimate resilient solutions. Outdated datasets can yield an ineffective LLM  against new threats, as affect all the later stages of incident response. This opens the door for new attacks and threats to be developed. As a consequence, ongoing research development is critically needed to develop resilient versions of LLMs with high ability to adapt to the changing landscapes.

\vspace{2mm} 
\noindent   
\textbf{Reflection of training datasets} 
Because cyber defense is one of the most important paradigms, building models for cyber threats detection is a very sensitive task \cite{razzaq2013cyber}. Training datasets of cyber defence techniques require specific criteria with wide range of cyber threats, cyber crimes scenarios, attacks patterns and defensive measures. On the other hand, LLMs can not train on traditional datasets of cyber security. As a consequence, there is not suitable datasets to build cyber defence specific LLMs. A need of clean dataset from any threats exploits, massive one, including cyber threats data and scenarios, providing smart attack patterns is critical.

\vspace{2mm} 
\noindent   
\textbf{Real-time suitability} 
On one hand, the rapid changes of cyber threats along with the sensitivity of cyber defence require the response of threat detection in short time \cite{brewer2015cyber}. On the other hand, LLMs require high computational resources and massive datasets for training and retraining. Both issues make the ability of LLMs to be integrated on a real-time mode is very limited. A special attention is needed to tackle this issue. Building efficient LLMs that are able to train on small datasets might be a direction to tackle that issue.

\vspace{2mm} 
\noindent   
\textbf{Using LLMs to develop attacks} 
The witnessed progression of LLMs on the machine learning paradigms presented promising solutions for all the realms of scientific research and applied sciences \cite{gordijn2023chatgpt}. It also impacted the development of cyber crimes and attacks. This power of recognizing the attacks and threats is accompanied by another potential advancement of cyber attacks. As a result, the landscape of cyber threats is growing wider and complicated. Early directions for organisations and governments to bound the development of GPT with legal standards and policies should be taken into account.   

\vspace{2mm} 
\noindent   
\textbf{Privacy Preserving} 
LLMs has a significant concern with privacy preserving, because their need to process massive amounts of datasets that include personal information. In other words, LLMs train on general purpose datasets such as Wikipedia, Glove and other dictionaries, therefore, personal information are easily accessible \cite{yan2024protecting}. These personal information may be reproduced from LLMs with simple cyber attacks. Data anonymization and implementing robust privacy-preserving techniques, such as differential privacy is crucial at this stage, however, it still challenging tasks. Using the power of LLMs with privacy preserving techniques remain an open challenge that requires special efforts from researchers and governments.

\vspace{2mm} 
\noindent   
\textbf{Adversarial attacks} 
LLMs can be exploited by introducing crafted inputs designed to deceive the models and then give wrong outputs. For example, some attacks can subtly alter the input data that result at incorrect, misleading, or harmful outputs \cite{yan2024protecting}. It is proved that LLMs are vulnerable to the adversarial attacks. Attackers might exploit this vulnerability to bypass security measures, manipulate automated decision-making processes. Developing robust LLMs against adversarial attacks will limit the power of LLMs to evolve, especially they do not have any processing on the input data. Adding methodology attacks to data attacks will make the challenge of LLMs more complicated. A possible direction here is to apply discriminative methodologies that are proves robustness on deep learning models such as \cite{hassanin2021mitigating} to LLMs. 

\vspace{2mm} 
\noindent   
\textbf{Interperability and transparency} 
The recent progression on GPT techniques paved the way to the involvement of AI in every branch in our daily life activities, including education \cite{moore2023empowering}, health \cite{sallam2023chatgpt}, driving \cite{chen2023driving}, etc. The result is that AI is about to dominate the decision making operations with a nature of "black box". With this given "black box" of LLMs, clear explanations are essential for trustworthiness and accountability. Also, security analysts should  understand how the final classification of a threat is made. The lack of transparency can hinder the adoption of LLMs. Traditional XAI models to be applied for LLMs is a probable direction \cite{zytek2024llms}.

Overall, these are the challenges in the field of cyber security and threat detection. We focused on mentioning the related challenges to cyber security to enable the researchers to concentrate on them, looking for quick solutions. Across the cyber security, there are more challenges such as internal hallucination, etc \cite{andriopoulos2023augmenting}.

% \section{Conclusion}
\section{Conclusions}
This paper surveyed close to 80 articles of LLMs within the cyber defense paradigm. Strengths and limitations of LLMs methods in cyber security are discussed. We cover the main branches of cyber defense including threat intelligence, vulnerability assessment, network security, privacy preservation, awareness and training, and ethical guidelines. A hierarchical structure for cyber defense sections is provided. 

Despite the power of the developed LLMs in overcoming a lot inherent issues in AI models, several challenges and open gaps still need to be filled, particularly when using these models in cyber defense. These challenges along research directions that are still open are listed and discussed. Although there are recent efforts proposed to address these limitations, it is still far from the acceptable solutions on such delicate field. This review will provide an overview picture for  researchers on open challenges and possible directions to enable developing better LLMs suited for cyber security applications.

% SlashNext Generative Human AI

%  BEC GAI Augmentation
%  Next Real-time Threat Detection Url Scanner
%  HumanAI leverages SlashNext’s LiveScan 
%  Sec-PaLM is based on PaLM 2
%  SentinelOne Purple AI
%  Slashnext Launches Industry’s First Generative AI Solution for Email Security,
%  Introducing Recorded Future AI: AI-driven Intelligence to Elevate Your Security Defenses, 
%  \textbf{BigID BigAI LLM}

% \input{sections/threatintelligence}
% \input{sections/networksecurity}

% \input{sections/pp}
% \input{sections/backup_intro}
% \input{sections/challenges}
% \input{sections/directions}
% \input{sections/conclusions}

\bibliographystyle{IEEEtran}
\bibliography{TNNLS}

\end{document}